 \newcommand{\y}{Y\-Ba$_{2}$\-Cu$_{3}$\-O$_{7-\delta}$}
 
 \newcommand{\bi}{Bi$_{2}$\-Sr$_{2}$\-Ca\-Cu$_{2}$\-O$_{8+\delta}$}
 \newcommand{\bidrei}{Bi$_{2}$\-Sr$_{2}$\-Ca$_{2}$\-Cu$_{3}$\-O$_{10+\delta}$}

\documentstyle[prb,aps,preprint]{revtex}  
 \tighten
 
\begin{document}
 \setcounter{equation}{0}  \draft             
\title{Detailed investigation of the superconducting transition of niobium 
   disks exhibiting the paramagnetic Meissner effect.}

\author{L.\ P\accent23ust, L.\ E.\ Wenger }
\address{Department of Physics and Astronomy, Wayne State University, 
              Detroit MI 48202 } 
\author{M.\ R. Koblischka}     
\address{SRL/ISTEC, 1-16-25 Shibaura, Minato-ku, Tokyo 105, Japan.} 
\date{\today}
\maketitle
\widetext
\begin{abstract}

The superconducting transition region in a Nb disk showing the 
paramagnetic Meissner effect (PME) has been investigated in detail. 
From the field-cooled magnetization behavior, two well-defined 
temperatures can be associated with the appearance of the PME: 
$T_{\rm 1 }\ (< T_{\rm c})$ 
indicates the characteristic temperature where the paramagnetic moment first 
appears and a lower temperature $T_{\rm p} \ (< T_{\rm 1})$ 
defines the temperature 
where the positive moment no longer increases. During the subsequent 
warming, the paramagnetic moment begins to decrease at $T_{\rm p}$ and then 
vanishes at $T_{\rm 1}$ with the magnitude of the magnetization change between 
these two temperatures being nearly the same as that during cooling. This 
indicates that the nature of the PME is reversible and not associated with 
flux motion. Furthermore, the appearance of this paramagnetic moment is 
even observable in fields as large as 0.2 T even though the magnetization 
does not remain positive to the lowest temperatures. Magnetic hysteresis 
loops in the temperature range between $T_{\rm 1}$ and $T_{\rm p}$ also exhibit 
a distinct shape that is different from the archetypal shape of a bulk type-II 
superconductor. These behaviors are discussed in terms of the so-called 
'giant vortex state'.

\end{abstract}

\narrowtext
%
%

A fundamental property of superconductivity is the Meissner effect, 
i.e., the occurrence of flux expulsion below the superconducting 
transition temperature, $T_{\rm c}$, and the resulting diamagnetic 
response to the applied magnetic field. In contrast to this 
behavior, several 
groups\cite{Svedlindh,Braunisch_PRL,Braunisch,Koetzler,KN,KA,Okram} 
have reported a 'paramagnetic' 
signal when cooling certain samples of high-$T_{\rm c}$ superconductors 
(mainly \bi\ ceramics, but also \bidrei\ powder,\cite{Braunisch} 
\y\ single crystals,\cite{KA} and Nd$_{2-x}$Ce$_x$CuO$_y$\cite{Okram}) in 
magnetic fields smaller than 0.1 mT through $T_{\rm c}$. This effect 
is now referred to as the paramagnetic Meissner effect (PME) or 
Wohlleben effect.

Several explanations for the origin of the PME have been given 
so far, including the formation of spontaneous currents due to 
$\pi$ contacts\cite{Braunisch} and spontaneous polarized orbital 
currents.\cite{orbital} Furthermore, it has been argued that 
this effect is a consequence of $d$-wave superconductivity.\cite{Rice} 
However, it still remains an open question why 
all high-$T_{\rm c}$ samples do not exhibit the PME.

The observation of a strikingly similar PME signal in disks 
of Nb,\cite{Detroit_PRL,Argonne} which is an isotropic 
$s$-wave (BCS) superconductor, revived the discussion about 
the origin of the PME.\cite{discussion} In turn, models based 
on flux trapping effects were proposed by Koshelev and Larkin (KL)\cite{KL} 
and Moshchalkov {\em et al.}\cite{Moshchalkov} 
Since the microstructure of all high-$T_{\rm c}$ samples exhibiting 
the PME is complicated by the presence of granularity, 
misorientation of grains,\cite{granular} intrinsic Josephson 
junctions,\cite{intrinsic} and the 2-dimensional character 
resulting from the layered crystalline structure,\cite{layered} 
Nb may serve as an ideal system to clarify the mechanism responsible 
for the formation of a paramagnetic moment during field-cooling for 
both conventional and high-$T_{\rm c}$ superconducting samples.

In order to provide a deeper insight into the transition regime of 
Nb, detailed magnetic investigations have been performed in the 
field-cooled cooling (FCC) and field-cooled warming (FCW) modes. 
The results of the present investigation suggest that the common 
picture of the PME has to be revised. The field-cooled magnetization 
results for Nb disks show a paramagnetic moment that has a 
{\em reversible} nature as it appears and disappears at the same 
temperature $T_{\rm 1}\ (<\ T_{\rm c})$ during cooling and warming, respectively. 
This moment is superimposed on a typical diamagnetic superconducting 
behavior below $T_{\rm 1}$ and continues to increase in magnitude until 
the temperature, $T_{\rm p}\ (<\ T_{\rm 1})$. Thus, the appearance of a net 
positive signal at lower temperatures depends on the relative magnitude 
of the paramagnetic moment of the sample to that of the diamagnetic 
moment below the temperature $T_{\rm p}$ .

Nb disks of diameter 6.4 mm were punched from 0.127-mm thick sheets 
(99.8 \% purity).\cite{14} Each disk was then positioned at the 
center of the second-order gradiometric detector coil in a commercial 
SQUID magnetometer\cite{15} with magnetic fields below 1 mT generated 
by a Cu solenoid which is part of its ultralow-field-option. To avoid 
any residual flux trapped in the surrounding superconducting magnet, 
the entire magnet system was warmed above its superconducting 
transition temperature and the superconducting magnet was never energized 
until the completion of all measurements below 1 mT. During both 
temperature and field measurements the sample was kept stationary, 
thereby eliminating any spurious signals that might have arisen from 
field inhomogenieties with sample position. The analog voltage output 
from the SQUID amplifier, which is proportional to the magnetic flux 
change through the pickup coils, was then recorded continuously during 
the temperature scan and transformed into magnetic moment values. 
A nominal temperature sweep rate of 35 mK/min for both cooling and 
the subsequent 
warming was achieved by direct low-level control of the heaters in the 
system using the EDC option.\cite{15} During magnetic-hysteresis-loop (MHL) 
measurements, the SQUID voltage component arising solely from the 
magnetic field sweep was subtracted from the measured voltages. The 
results presented in this communication are for one particular Nb disk 
(NbD4S2), see also Ref.\ \onlinecite{Detroit_PRL}.

Figure\ \ref{F1} presents detailed magnetic moment $m(T)$ results 
in the superconducting transition region for field-cooled cooling 
(FCC) and field-cooled warming (FCW) modes in magnetic fields 
between 10 and 200 $\mu$T. The temperature was continuously swept 
from 9.4 K down to 8.5 K and 
then the temperature sweep direction is reversed. 
These scans of the magnetic moment, $m(T)$, reveal several 
characteristic temperatures. $T_{\rm c}$ defines the onset of 
superconductivity and of a diamagnetic moment, which occurs at 
$\approx$ 9.23 K as shown in the inset of Fig.\ \ref{F1}. Below 
$T_{\rm c}$, $m(T)$ becomes more negative upon cooling until 
$T_{\rm 1}$ ($\approx$ 9.16 K at the lowest fields) where $m(T)$ 
abruptly turns towards positive values. This increase in $m(T)$ 
continues until another characteristic temperature, $T_{\rm p}$ 
($\approx$ 
9.16 K at the lowest fields), is reached where $m(T)$ exhibits a 
cusplike behavior. Below $T_{\rm p}$, the relative temperature 
dependence of $m(T)$ appears to correspond to the archetypal FCC 
behavior of a superconducting sample not exhibiting the PME. As 
discussed in Ref.\ \onlinecite{Detroit_PRL}, another characteristic 
feature of the PME for the Nb disks is the large hysteresis between 
the FCC and FCW curves with the FCW curves being lower than the FCC 
curves. Upon warming, the $m(T)$ curves remain practically constant 
until the moment turns abruptly towards negative values at a 
temperature essentially the same as $T_{\rm p}$. Then, at a 
temperature identical to $T_{\rm 1}$, $m(T)$ jumps to a less 
diamagnetic and more probable equilibrium value as 
the PME appears to suddenly vanish. The similarity in characteristic 
temperatures where the positive moment first appears during cooling 
and then disappears during warming as well as where the positive 
contribution stops increasing in magnitude during cooling and 
subsequently begins to decrease upon warming suggests that the PME 
has a reversible nature.

The same basic features can even be observed in the FCC and FCW 
curves recorded at magnetic fields as high as 0.2 T. In 
Fig.\ \ref{F2} the $m(T)$ behaviors are presented for applied fields 
from 20 to 85 mT. The onset of diamagnetism just below $T_{\rm c}$ 
in the FCC curves is still present as illustrated in the inset for 
a field of 85 mT. The onset of a diamagnetic moment at $T_{\rm c}=$ 
8.6 K is followed by a very sharp positive rise in the FCC curve 
at $T_{\rm 1}$ = 8.32 K which eventually reaches a maximum at 
$T_{\rm p}$. However, instead of remaining positive for all 
temperatures below $T_{\rm p}$ as in the low-field measurements, 
the overall $m(T)$ becomes diamagnetic which suggests that the 
diamagnetic component outweighs the paramagnetic component at these 
higher fields. Nevertheless, the paramagnetic component is still 
present in the sample at $T \ll T_{\rm p}$ as evidenced by the 
corresponding FCW curves. Even though $m(T)$ during FCW is always 
diamagnetic up to $T_{\rm c}$, there is a noticeable change towards 
larger diamagnetic values at a temperature very close to $T_{\rm p}$ 
followed by a jump towards a vanishingly small $m(T)$ value at a 
temperature similar to $T_{\rm 1}$ (see inset). With decreasing 
magnetic field, this jump becomes increasing less broad and eventually 
becomes discontinuous at the lowest fields. This characteristic 
behavior associated with the appearance and disappearance of the 
positive moment even in high fields gives further credence to the 
reversible nature of the PME. 

Figure\ \ref{F3} further illustrates the reversible nature of the 
paramagnetic component of the magnetic moment. While the field 
is kept fixed at $\mu_o H_{\rm ext}$ = 50 $\mu$T, the temperature 
is cooled to 
various temperatures $T_{\rm min} (< T_{\rm p})$ and then subsequently 
raised above $T_{\rm c}$. While all FCC curves overlap as expected, 
the FCW curves below $T_{\rm p}$ are nearly temperature independent 
with the magnitude being dependent upon $T_{\rm min}$. Above 
$T_{\rm p}$, $m(T)$ during FCW decreases rapidly until $T_{\rm 1}$.  
Just below $T_{\rm 1}$ $m(T)$ jumps abruptly 
towards a value close that of the FCC 
curve suggesting that the FCC curves above $T_{\rm 1}$ are in 
equilibrium. The size of this jump at $T = T_{\rm 1}$ depends only 
on the moment value at the onset of the step at $T_{\rm p}$. 
Further note that the shape and size of the change in $m(T)$ between 
$T_{\rm p}$ and $T_{\rm 1}$ are nearly the same for all warming 
curves with each $m(T)$ curve being shifted downwards for lower 
$T_{\rm min}$. Even the complementary FCC curves appear to have 
a similar shape and magnitude change in this temperature region 
suggesting the paramagnetic component is non-hysteretic. 

To estimate 
the separate paramagnetic and diamagnetic components, the PME 
contribution is assumed to be constant below $T_{\rm p}$. The 
resulting FCC and FCW data below $T_{\rm p}$ are then shifted by 
$m_{\rm shift} =$ 1.33 $\times$ 10$^{-9}$ Am$^2$ as shown in 
Fig.\ \ref{F3} for the $T_{\rm min}$ = 8.2 K data. The resulting 
'shifted' curves with the interpolations between $T_{\rm 1}$ and 
$T_{\rm p}$ (dashed lines) appear to form FCC and FCW curves for 
a archetypal type-II superconductor. Thus the characteristic 
$m(T)$ behavior can be regarded as consisting of two additive moments: 
a hysteretic behavior arising from the flux trapping and diamagnetic 
screening associated with a typical type-II superconducting sample 
and a {\em nearly reversible} behavior associated with the PME of 
the sample which appears and disappears at a well-defined temperature 
of $T_{\rm 1}$. At low-fields, the aburpt appearance (disappearance) 
of this positive moment upon cooling below $T_{\rm 1}$ (warming just 
below $T_{\rm 1}$) is fairly spontaneous similar to the onset 
(disappearance) of global diamagnetic screening currents at 
$T_{\rm c}$ rather than the viscous nature exemplified by flux flow. 
Below $T_p$, this moment apparently does not change with temperature 
and can be regarded as an additive constant to the field-cooled 
magnetic moment of a non-PME superconductor. 

To gain further insight into the nature of this additive paramagnetic 
moment for the Nb disks, magnetic hysteresis loops (MHLs) were measured. 
A set of MHLs recorded in the temperature range from 9.04 to 9.09 K, 
i.\ e., around $T_{\rm p}$, are presented in Fig.\ \ref{F4}. The 
MHLs between $T_{\rm 1}$ and $T_{\rm p}$ are found to exhibit several 
distinct features which probably have the same physical origin as the 
appearance of the paramagnetic moment during cooling in small applied 
fields. Between $T_{\rm c}$ and $T_{\rm 1}$, the sample is 
superconducting, but only a small diamagnetic screening current can be 
induced by ramping 
$H_{\rm ext}$. Below $T_{\rm 1}$, the magnitude of the critical 
currents increases significantly, and the MHLs exhibit a strange, 
nearly parallelogram-like shape. Note that the reverse legs are 
entirely linear, all 'corners' are quite sharp, and the size of the 
parallelograms increases with decreasing temperature. Also, the shape 
of the initial curves is very linear and merges with the loop quite 
abruptly. Just below $T_{\rm p}$, a remarkable change in the shape of 
the MHLs takes place as the MHLs acquire the archetypal behavior for 
a bulk type-II superconductor with a maximum close to $H_{\rm ext} = 0$. 
Additionally, the slope ${\rm d}m/{\rm d}H_{\rm ext}$ of the initial 
curves changes by about 20 \% below $T_{\rm p}$ with the magnitude of 
this change corresponding to the step observed in the zero field-cooled 
magnetization curves.\cite{Detroit_PRL} 

As reported previously, the sample surface plays a crucial role in 
the appearance of the paramagnetic moment in the Nb disk sample. 
These PME effects were found to vanish after mechanical abrading 
the disk's top and bottom surfaces and reducing the overall thickness 
by about 10 \%\cite{Detroit_PRL,Argonne} while a net positive magnetic 
moment was induced into a thicker Nb disk by means of ion implanting.\cite{Detroit_2} 
Neither the simple approach of flux compression by KL\cite{KL} nor the 
incorporation of the surface effects\cite{Argonne} 
seem to explain the features 
reported here. However, our measurements appear to be more consistent 
with a superconducting current forming a large 'giant vortex' on the 
sample surface. Giant vortex states were first described theoretically 
by Fink and Presson\cite{Fink} and more recently as a possible source 
for the PME by Moshchalkov {\em et al.}.\cite{Moshchalkov} 
The basic premise is that the giant vortex state with a fixed orbital 
quantum number is formed as the superconductor is field-cooled at the 
third (surface) critical field $H_{\rm c3}(T)$. With decreasing 
temperature, the superconducting order parameter would grow and compress 
the trapped flux into a giant vortex leading to a positive or paramagnetic 
response. However, the original model pictured a single giant vortex 
developing from the edges of a cylindrical sample as compared to several 
individual vortices developing at microstructural defects on the disk's 
surface. Thus it is important to determine the flux and/or screening 
current distributions in samples exhibiting the PME in the transition 
region between $T_{\rm 1}$ and $T_{\rm p}$. Using highly sensitive imaging 
techniques,\cite{imaging} e.\ g., could clarify the origin of the PME in Nb 
as well as elucidate our understanding of the PME in high-$T_{\rm c}$ 
superconductors. In granular samples, the surface of the individual grains 
may have deteriorated thus providing a similar surface layer with different 
superconducting properties as on our Nb disk surfaces.

In summary, the present measurements suggest that the nature of the PME is 
reversible and not associated with flux motion as the total magnetization 
consists of two components: a nearly reversible positive moment superimposed 
on a hysteretic diamagnetic superconducting behavior below $T_{\rm 1}$. 
Consequently, the appearance of a net positive signal at lower temperatures 
depends on the relative magnitude of the diamagnetic moment of the sample 
to that of the paramagnetic moment.

L.\ P.\ and L.\ E.\ W.\ acknowledge support from NATO Grant No.\ 961357 while 
M.\ R.\ K.\ is supported by STA.

\begin{figure}
\caption[]{ Low-field field cooled $m(T)$ curves recorded during cooling 
      (FCC - open symbols) and warming (FCW - solid symbols) on sample 
      NbD4S2 in various 
      fields $\mu_o H_{\rm ext} =  10 \mu$T (circles), 50 $\mu$T  
      (triangles), 100 $\mu$T (diamonds), and 200 $\mu$T (squares). 
      The sharp, downward turn in the $m(T)$ curve  
      corresponds to $T_{\rm p}=$ 9.05 K. \\
      {\em Inset}: Details of FCC curves just below the critical 
      temperature indicating sharp 
      turn towards positive values at temperature $T_{\rm 1}$. 
     $T_{\rm 1}$ at various fields is indicated by a dashed line.}
\label{F1}
\end{figure}

\begin{figure}
\caption[]{ FCC and FCW $m(T)$ curves recorded on sample NbD4S2 in fields 
      $\mu_o H_{\rm ext}=$ 20, 60 and 85 mT. Note the significant step on 
      the FCC curve towards positive values and the complementary step 
      towards negative values on the FCW curve. \\
      {\em Inset}: Details of the curves at 85 mT with  
      onset of diamagnetic moment at 8.6 K followed by a very sharp 
      onset of a positive step on the FCC curve at $T_{\rm 1}=$ 8.32 K. }
\label{F2}
\end{figure}

\begin{figure}
\caption[]{ FCC and FCW $m(T)$ curves for a constant field $\mu_o H_{\rm ext} =$ 50 $\mu$T 
      measured down to various $T_{\rm min} < T_{\rm p}$.  
      To estimate the magnitude of the paramagnetic step on the FCC and FCW curves, 
      both curves below $T_{\rm p}$ are shifted by $m_{\rm shift}=$ 1.33 $\times$ 10$^{-9}$ 
      Am$^2$ (dashed arrows) to such a position that $m(T)$ above 
      $T_{\rm 1}$ and below $T_{\rm p}$ point toward each other. \\
      {\em Inset}: Details of FCC and FCW curves around $T_{\rm 1}$. }
\label{F3}
\end{figure}

\begin{figure}
\caption[]{  Magnetic hysteresis loops (MHL) in the temperature range 9.04 $\leq T \leq$  
      9.09 K for 0.01 K increments. Note the remarkable change of the MHL shape between 
      9.05 and 9.06 K, i.e., at $T=T_{\rm p}$. }
\label{F4}
\end{figure}

\end{document}